\documentstyle[epsf]{mn}  

\begin{document}   
\voffset -0.3truecm
\title[Effects of the UV background radiation on galaxy formation]
{\vglue -2.0truecm{
\rightline{\Large
}
\vglue -0.25truecm
{\rightline{{\Large\bf OU-TAP 78}}}
\vglue 1.5truecm}
\vglue -2.2truecm{
\rightline{\Large
}
\vglue -0.25truecm
{\rightline{{\Large\bf KUNS-1513}}}
\vglue 1.5truecm}
\noindent 
{Effects of the UV background radiation on galaxy formation}
\author[M. Nagashima, N. Gouda \& N. Sugiura]{Masahiro Nagashima,$^{1}$
Naoteru Gouda$^{1}$ and Norimasa Sugiura$^{2}$\\
$^{1}$Department of Earth and Space Science, Graduate School of Science,   
Osaka University, Toyonaka, Osaka 560-0043, Japan\\    
$^{2}$Department of Physics, Kyoto University, Sakyo-ku, Kyoto
606-8502, Japan\\ 
}}
\maketitle   

\begin{abstract}   
  We investigate the effects of the UV background radiation on galaxy
  formation by using the semi-analytic model including the
  photoionization process.  The semi-analytic model is based on Cole
  et al. and we use almost the same parameters as their `fiducial'
  model.  We find that the UV background mainly affects the formation
  of dwarf galaxies when $J_{-21}\ga 1 (J\equiv J_{-21}\times
  10^{-21}$ ergs cm$^{-2}$ s$^{-1}$ sr$^{-1}$ Hz$^{-1}, J$ is the
  intensity of the UV background).  Because of the suppression of star
  formation, the number density of small objects corresponding to
  dwarf galaxies decreases compared with the case of no UV radiation
  when the UV background exists until the present epoch.  On the
  contrary, the UV radiation hardly affects massive galaxies.  This is
  because the massive galaxies are formed by mergers of small
  galaxies, which are formed at high redshift when the effect of the
  UV background is negligible.  This strongly suggests that it is
  important to consider the merging histories of the galaxies.  On the
  other hand, when the UV background vanishes at a low redshift
  ($z\sim 2$), the number density of small objects is hardly changed
  but the colour becomes bluer, compared with the case of no UV
  radiation, because stars are newly formed after the UV background
  vanishes.
  
  Moreover, we show the redshift evolutions of the luminosity
  functions and the colour distributions of galaxies.  Because the
  effect of the UV background is strong at low redshift, we can
  discriminate between the types of evolution of the UV background by
  observing the evolution of the luminosity function and the colour
  distributions, if the UV intensity is sufficiently strong
  ($J_{-21}\ga 1$).
\end{abstract}   
   

\section{INTRODUCTION}   
The problem of galaxy formation is one of the most important
unresolved problems in astrophysics.  Recently, owing to the
surprising development of the technology concerned with telescopes
(e.g. the {\it Hubble Space Telescope} ({\it HST}) and the Keck
telescope), we can see the birth and evolution of galaxies.  Thus we
can obtain some significant clues to understanding the galaxy
formation processes.

It is known that the intergalactic medium (IGM) is highly ionized
through the Gunn--Peterson test (Gunn \& Peterson 1965).  Ultraviolet
(UV) background radiation, which ionizes the IGM, also affects the
galaxy formation processes by photoionizing gas clouds.  The UV
photons penetrate the gas clouds and heat up the gas, then star
formation in the gas cloud is suppressed by the UV radiation.
Therefore it is important to take into account the effects of the UV
radiation when we consider the galaxy formation processes.

The sources of the UV background radiation are considered to be mainly
quasi-stellar objects (QSOs).  From the observations of Ly{$\alpha$}
clouds and QSOs, the intensity of the UV background radiation is
indirectly measured by the `proximity effect' (Carswell et al. 1984;
Bajtlik, Duncan \& Ostriker 1988; Bechtold 1994; Kulkarni \& Fall
1993; Williger et al. 1994; Giallongo et al. 1996; Lu et al. 1996).
The evolution of the UV intensity at low redshifts ($z\la 2$) is
approximately proportional to $(1+z)^{\gamma}$, $\gamma\sim 4$ from
the evolution of the QSO luminosity function (Pei 1995), while at high
redshifts ($z\ga 2$) the evolution of the UV background is not yet
determined.

It has been pointed out that the UV background affects the galaxy
formation (e.g. Dekel \& Rees 1987).  Babul \& Rees (1992) and
Efstathiou (1992) have discussed the suppression of star formation by
the UV background in dwarf galaxies by considering the evolution of a
single galactic gas cloud.  Taking into account the photoionization
process, the mass function of cooled objects has been calculated in a
hierarchical clustering scenario by Chiba \& Nath (1994).  Their
result is that the number density of massive galaxies decreases
compared with the case of no UV background.  This behavior is
explained as follows.  In the hierarchical clustering scenario, large
clouds generally collapse later than small clouds.  As the mean
density of a cloud is assumed to be about 200 times the cosmological
background density at the collapsing epoch (see, e.g., Peebles 1993),
large clouds have a lower density than small clouds.  Then the UV
photons penetrate deeply into the large clouds and suppress the star
formation, which leads to the decrease in the number density of the
massive galaxies.  However, mergers of galaxies actually occur in the
hierarchical clustering model.  If luminous galaxies are formed by the
mergers of faint galaxies that could collapse at high redshift with
little effects of the UV radiation, the number of such luminous
galaxies are probably less affected by the UV background.  Therefore,
it is important to consider the merging history of galaxies in order
to clarify the effects of the UV radiation.

The semi-analytic approach to the galaxy formation, which includes the
merging history of dark haloes and galaxies, has been developed
recently (e.g. Kauffmann, White \& Guiderdoni 1993; Cole et al. 1994;
Roukema et al. 1997; Somerville \& Primack 1998).  It has already been
confirmed that the averaged stellar age of each galaxy has only a weak
correlation with the luminosity of each galaxy by the semi-analytic
model (Kauffmann \& Charlot 1998).  This result would be inconsistent
with the shape of the cold dark matter (CDM) mass spectrum unless the
mergers occur.  This fact also motivates us to analyse the effects of
the UV background on galaxy formation via the semi-analytic approach.

In this paper, we investigate the effects of the UV background
radiation with the semi-analytic model, which is an extension of Cole
et al. (1994) based on the block model for merging histories of dark
haloes (Cole \& Kaiser 1988; Nagashima \& Gouda 1997) by including the
photoionization process.  We will show how the luminosity functions
and the colour distributions of galaxies are changed by the UV
background.

In Section 2, we describe our models.  In Section 3, we show the
luminosity function and the colour distribution of model galaxies, and
their redshift evolution.  Section 4 is devoted to conclusions and
discussion.

\section{MODELS}
\subsection{Semi-analytic model}
We use the semi-analytic model, originally developed by Cole et al.
(1994).  The model includes the following processes: merging histories
of dark haloes, gas cooling and heating, star formation and feedback,
mergers of galaxies, and stellar population synthesis.  We include the
effect of the UV background radiation in their original model.  The
parameters that we use are the same as the `fiducial' model in Cole et
al. (1994), except for the block masses and the stellar population
model.  The astrophysical parameters in the fiducial model are chosen
by matching the bright end of the observed $B$-band luminosity
function.  The cosmological parameters are specified as follows:
$\Omega_{0}=1, \Lambda_{0}=0, H_{0}=50$ km s$^{-1}$ Mpc$^{-1}$,
$\sigma_{8}=0.67$ and $\Omega_{b}=0.06$.  We use two maximum block
masses $M_{0}$ corresponding to $2\times 10^{16}h^{-1}$M$_{\odot}$ and
$\sqrt{2}\times 10^{16}h^{-1}$M$_{\odot}$ in order to avoid the
influence of the artificial discreteness of masses of blocks,
$M_{i}=M_{0}/2^{i}$.  The number of the hierarchy is 24.  We use the
GISSEL96 as the stellar population model provided by Bruzual \&
Charlot (in preparation).  More details about the original model are
found in Cole et al. (1994).  The data we will show are averaged over
30 realizations by using each maximum block mass.

\subsection{Photoionization and mass fraction of neutral core}
The UV radiation permeates a gas cloud to a radius at which the
ionization by the UV photons balances with the recombination of ions
and electrons.  For simplicity, we assume that the gas within the
`cooling radius' $r_{cool}$, at which the cooling time scale without
UV radiation equals the lifetime of the halo including the gas, cools
and recombines rapidly.  The lifetime of the halo is defined as the
time between the collapse of the halo and its subsequent incorporation
into a larger halo.  In Cole et al. (1994), the cooled gas within
$r_{cool}$ is interpreted as the star-forming gas.  When the UV
background exists, however, the mass of the star-forming gas decreases
by the penetration of the UV background.  Then we estimate a radius
$r_{UV}$ at which the UV photons are perfectly absorbed.  We assume
that the intensity of the UV radiation begins to decrease at
$r_{cool}$ and vanishes at $r_{UV}$ by absorption of the UV photons
($r_{UV}\leq r_{cool}$), and that the gas within $r_{UV}$ is cooled
and forms stars.  With the `inverted' Str{\"o}mgren sphere
approximation, we obtain this radius $r_{UV}$ by the following
relation,
\begin{equation}
\int_{r_{UV}}^{r_{cool}}n_{p}(r)n_{e}(r)\alpha^{(2)}(T_{eq})4\pi
r^{2}dr =\pi (4\pi r_{cool}^{2})1.5\times 10^{5}J_{-21},
\end{equation}
where $n_{p}$ and $n_{e}$ are the proton and electron number
densities, respectively, $\alpha^{(2)}(T_{eq})$ is the recombination
coefficient to all excited levels at a temperature $T_{eq}$, and
$J_{-21}$ is the normalized intensity of the UV radiation, $J\equiv
J_{-21}\times 10^{-21}$ ergs cm$^{-2}$ s$^{-1}$ sr$^{-1}$ Hz$^{-1}$.
As the density distribution of the gas is assumed to be an isothermal
distribution ($\propto r^{-2}$), the above equation is solved as
follows:
\begin{equation}
r_{UV}=r_{cool}\left[1+\left(\frac{r_{cool}}{R}\right)^{3}
\frac{A(J_{-21},T_{eq},M)}{(1+z)^{5}}\right]^{-1},\label{eqn:uv}
\end{equation}
where
\begin{equation}
A(J_{-21},T_{eq},M)\equiv\frac{1.35\times 10^{6}\pi
J_{-21}}{\alpha^{(2)}(T_{eq})n_{R}^{2}R(M)},\label{eqn:defa}
\end{equation}
where $R(M)$ is the virial radius of the dark halo with the mass $M$
in the comoving coordinate, and $n_{R}$ is the mean comoving baryon
density within $R(M)$, which is about 200 times the cosmological
background baryon density under the assumption of the spherical
collapse.

In Fig.1, we plot the ratio $r_{UV}/r_{cool}$ from the above function
at various redshifts in the case of $J_{-21}=0.1$, $R(M)=r_{cool}$ and
$M_{baryon}=0.06M$, where $M_{baryon}$ is the baryon gas mass.  Note
that the curve at $z=0$ is in agreement with the horizontal axis
because UV photons penetrate into the vicinities of the centres of gas
clouds.  In our calculations shown in Section 3, $r_{cool}$ is a
function of the lifetime of haloes, which depends on their merging
histories.  It is assumed that $T_{eq}=3\times 10^{4}$ K.  As the mean
density of clouds at the collapsing epoch, $n_{R}$, is equal to
$200n_{b}$, where $n_{b}$ is the cosmological background density,
clouds are denser at high redshifts and so the UV photons are
perfectly absorbed at outer region.  Therefore there is little effect
of the UV radiation at high redshifts.  At all epochs, the smaller gas
cloud is, the smaller the ratio $r_{UV}/r_{cool}$ becomes, because the
effect of the photoionization is enhanced in small clouds by
$A(J_{-21},T_{eq},M)\propto J_{-21}R(M)^{-1}\propto J_{-21}M^{-1/3}$.
Note that $n_{R}\simeq 200n_{b}$ is common to all mass objects that
are just collapsed at an epoch.  We also show the cases of
$J_{-21}=0.01$ and $1$ in order to see the effect of changing the
intensity.  It is found that the ratio $r_{UV}/r_{cool}$ increases
when $J_{-21}$ decreases.  As the second term in the brackets of
equation (\ref{eqn:uv}) scales as $J_{-21}(1+z)^{-5}$, increasing
$J_{-21}$ corresponds to decreasing $z$.

\begin{figure}
\epsfxsize=8cm
\epsfbox{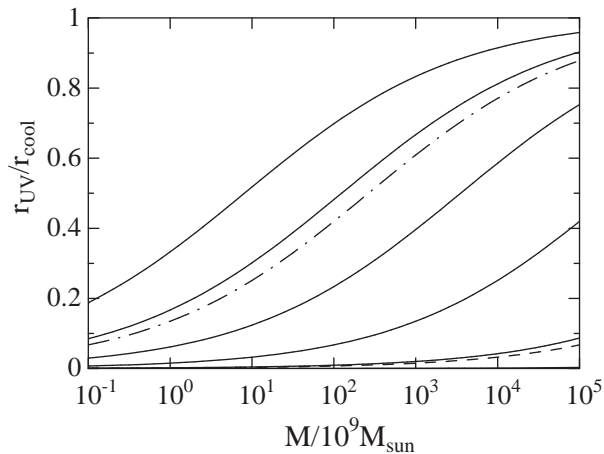}
\caption{Mass fraction of neutral core.  The horizontal axis shows the
  total mass including dark matter.  The solid lines denote the ratio
  $r_{UV}/r_{cool}$ in the case of $J_{-21}=0.1$ at $z=5, 4, 3, 2, 1$
  and $0$ from upper to lower, respectively.  Note that the curve at
  $z=0$ is almost the same as the horizontal axis.  The dash-dotted
  line and the dashed line show the ratio in the case of
  $J_{-21}=0.01$ and $1$ at $z=2$, respectively.}
\end{figure}

It is reported that this inverted Str{\"o}mgren sphere approximation
is not so good when we solve the radiation transfer dependently on the
frequency of the photons (Tajiri \& Umemura 1998).  They have found
that the UV photons permeate the clouds far beyond the radius at the
optical depth $\tau\sim 1$.  In our calculations, we have also used a
criterion that has the same dependence on mass and UV intensity as the
criterion by Tajiri \& Umemura (1998) as a trial.  We define the
critical density $n_{crit}$ as the density corresponding to the radius
$r_{UV}$ at which UV photons are perfectly absorbed, as follows:
\begin{equation}
n_{crit}\simeq
10^{-2}\mbox{cm}^{-3}\left(\frac{M}{10^{8}M_{\odot}}\right)^{-1/5}
J_{-21}^{3/5}.\label{eqn:tu}
\end{equation}
This critical density is almost the same as given by Tajiri \& Umemura
(1998).  We have found that the results are qualitatively the same as
those by using the inverted Str{\"o}mgren sphere approximation.
However, they assume the homogeneous density cloud instead of the
isothermal sphere exposed to the UV radiation.  Then we cannot obtain
the exact criterion for the isothermal sphere from their work.

Moreover, $r_{UV}$ may become much smaller if we consider the
condition under which H$_{2}$ molecules can be formed.  It is
inadequate for forming stars that the gas temperature becomes
$10^{4}$K (Susa, private communication).

Anyway, it is difficult at present to fix a reasonable criterion for
star formation under the photoionization effect of UV radiation.
Thus, we will proceed as follows.  First, note that assuming smaller
$r_{UV}$ corresponds to increasing intensity of $J_{-21}$. Then, as a
trial, we investigate the effect of the photoionization by changing
the value of $J_{-21}$ instead of changing the criterion of star
formation.  Hereafter we study the cases $J_{-21}=0.1$ ({\it weak
condition}) and $J_{-21}=10$ ({\it strong condition}).

\subsection{UV background radiation}
There are many uncertainties in the determination of the evolution of
the UV background, although it is suggested that the intensity of the
UV background rapidly declines at low redshifts ($z\la 2$) as
mentioned in the Introduction.  In order to take into account possible
evolutions of the UV background, we calculate the following cases: (1)
no UV background exists, (2) $J_{-21}=0.1$ or $10$ for $0\leq z\leq
5$, (3) $J_{-21}=0.1$ or $10$ for $2\leq z\leq 5$, and (4) $J\propto
(1+z)^{\gamma}$, where $\gamma=4$ for $z\leq 2$ and $\gamma=-1$ for
$z\geq 2$, and $J_{-21}=0.1$ or $10$ at $z=2$.  Case (4) is probably
the most realistic of these four cases.  We show these models in
Fig.2.

\begin{figure}
\epsfxsize=8cm
\epsfbox{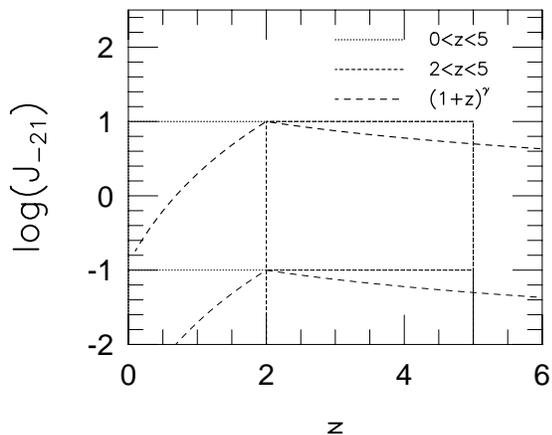}
\caption{Evolution of UV background intensity.  The thick and
  thin lines denote the strong and weak condition, respectively.  The
  dotted lines, short-dashed lines and long-dashed lines show cases
  (2) UV radiation exists in $0\leq z\leq 5$, (3) $2\leq z\leq 5$, and
  (4) $J\propto (1+z)^{\gamma}, \gamma=4$ for $z\leq 2$ and
  $\gamma=-1$ for $z\geq 2$.}
\end{figure}

\section{RESULTS}
\subsection{Luminosity function}
\subsubsection{Weak condition}
In Figs.3(a) and 3(b), we show the $B$- and $K$-band luminosity
functions of model galaxies, respectively.  The solid lines, dotted
lines, short-dashed lines and long-dashed lines show cases (1), (2),
(3) and (4) for the weak condition.  The $B$-band data are taken from
Loveday et al. (1992) and converted from their observed $b_{J}$ to
Johnson B by $B=b_{J}+0.2$.  The K--band data are those from Mobasher,
Sharples \& Ellis (1993).  Because these data have also been used by
Cole et al. (1994), a direct comparison of our results with the
results in Cole et al. (1994) becomes easier.

We also show the redshift evolution of the $B$-band luminosity
functions in the galaxies' rest frame in Figs.4.  The types of lines
are the same as Figs.3.  The $K$-band luminosity functions are not
shown, because their properties of the redshift evolution are almost
the same as those of the $B$-band luminosity functions.  The output
redshifts are (a) $z=0$, (b) $z=0.4$, (c) $z=1$, and (d) $z=2$.  At
$z=2$, the lines for case (2) are in agreement with those for case (3)
by their definitions.

The differences between luminosity functions for various cases become
small at high redshifts.  This is considered as follows.  At high
redshifts $z\ga 2$, mergers of galaxies occur frequently, so the
lifetime $t_{life}$ becomes very short.  This leads to the small
cooling radius $r_{cool}$, and so this effect results in the same
effect as the reduction of the UV intensity, because the effect of the
photoionization emerges via the form $(r_{cool}/R)^{3}J_{-21}$ from
equation (\ref{eqn:uv}).  Therefore, such merging galaxies are
slightly affected by the UV background at $z\ga 2$.

We find that the UV background hardly affects the galaxy formation in
the weak condition.  The number density of faint objects decreases
slightly in cases (2) and (4), but these differences from case (1) are
within the observational errors.

As shown in Fig.3, the slight differences between the luminosity
functions in the different UV evolution models in the weak condition
arise only in the faint end, because the UV background affects small
galaxies more strongly.  These differences become smaller at higher
redshifts (Fig.4), and are smaller than the observational errors.
Thus we cannot distinguish characteristics of the evolution of the UV
background by observing the present luminosity functions when the UV
intensity is weak.

\begin{figure}
\epsfxsize=9cm
\epsfbox{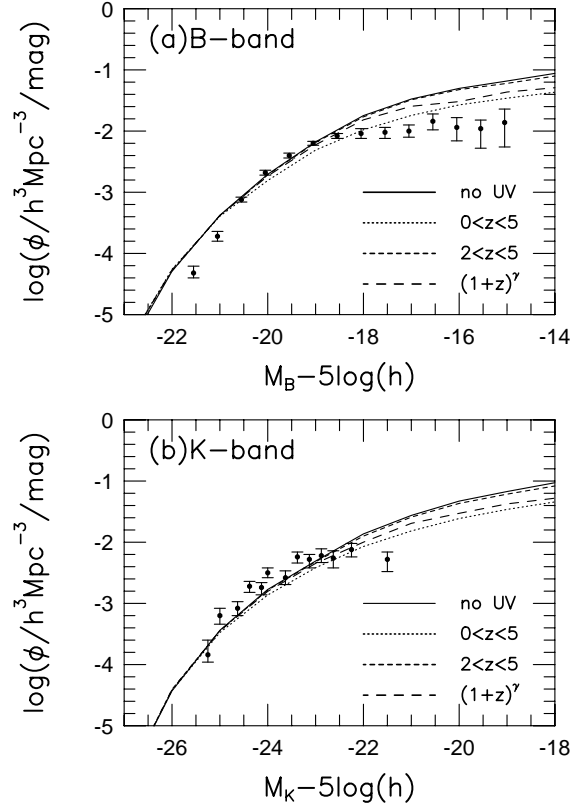}
\caption{(a) $B$-band luminosity functions.  (b) $K$-band luminosity
  functions.  The solid lines, dotted lines, short-dashed lines and
  long-dashed lines show cases (1) no UV, (2) UV radiation exists in
  $0\leq z\leq 5$, (3) $2\leq z\leq 5$, and (4) $J\propto
  (1+z)^{\gamma}, \gamma=4$ for $z\leq 2$ and $\gamma=-1$ for $z\geq
  2$.  In all cases the weak condition is used.  The filled circles
  with errorbars denote the luminosity functions from (a) Loveday et
  al. (1992) and (b) Mobasher et al. (1993).}
\end{figure}
\begin{figure}
\epsfxsize=7cm 
\epsfbox{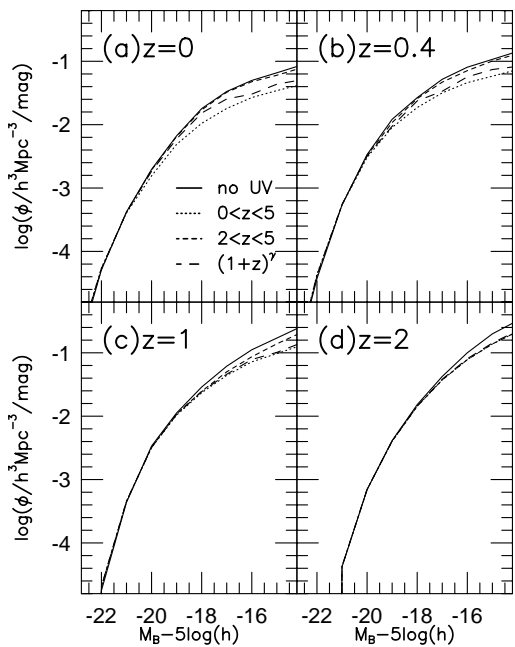}
\caption{$B$-band luminosity functions in galaxies' rest frame at
  various redshifts. (a) z=0, (b) z=0.4, (c) z=1, and (d) z=2.  The
  types of lines are the same as Fig.3.  In all cases the weak
  condition is used.}
\end{figure}

\subsubsection{Strong condition}
In the strong condition, the differences between luminosity functions
for various cases are larger than the observational errors (Figs.5 and
6).  Here, we see the properties of luminosity functions for various
cases of the UV background in the strong condition.

\paragraph{Merger effect of galaxies}
In order to investigate the effect of galaxy mergers, we compare the
luminosity functions for case (1) with those for case (2).  In Fig.5,
we show the same figures as Fig.3, but for the strong condition.  The
number of faint objects ($M_{B}-5\log(h)\simeq -18$) decreases about a
factor of 7.5 in case (2), compared with case (1).  On the other hand,
the most luminous galaxies are less affected by the UV background,
although their number slightly decreases.

In Chiba \& Nath (1994), the characteristics of luminosity (mass)
functions when the UV background exists are opposite to ours, i.e.
the number of dwarf galaxies is not affected by the UV background,
while the formation of massive galaxies is much suppressed in their
results.  This is because in our calculation the luminous galaxies are
formed in high density regions and formed by mergers of small galaxies
formed at early epochs owing to the high density environments, at
which there is only a little effect of the UV background because of
the high density of gas clouds (see Fig.1).  On the other hand, as
most of the dwarf-scale galaxies at high redshifts are merged into
luminous galaxies until the present epoch, dwarf galaxies at present
are formed at low redshifts compared with the typical epoch estimated
from the power spectrum of density fluctuations, so that such dwarf
galaxies are much affected by the UV background and their number
decreases.

In Fig.6, at all redshifts, the number of faint galaxies is small in
case (2); besides, the number of luminous galaxies is also small at
the low redshifts.  This is because the number of progenitors of such
luminous galaxies at high redshift is smaller than that in case (1).
Therefore, even the most luminous galaxies become slightly fainter.
Thus the number of the most luminous galaxies seems to decrease
slightly.

\paragraph{Delayed formation of galaxies.}
Here we consider case (3).  When the UV radiation ceases at $z=2$, the
number of galaxies at present does not change compared with the case
of no UV radiation.  Therefore if the UV background vanishes at $z=2$,
we cannot distinguish the types of the evolution of the UV background
from the luminosity function at $z=0$ (Fig.5).

After the UV background vanishes at z=2 in case (3), the number of
galaxies increases rapidly and its luminosity function shows good
agreement with case (1) at $z\la 0.4$ (Fig.6).  This increase occurs
similarly over the entire range of magnitudes.  This means that the
epoch of galaxy formation is delayed as a whole by the UV background
(Babul \& Rees 1992).

\paragraph{Effect of mass dependence of photoionization}
Now we consider case (4).  In this case, the intensity of the UV
background evolves continuously, so that the epoch when clouds become
free from the effect of the UV background at low redshifts depends on
their mass, and the epoch when massive clouds become free from the UV
background is earlier than that of small clouds (see Fig.1).

It is interesting that the number density of faint galaxies in case
(4) is in agreement with that in case (2), while the number density of
luminous galaxies in case (4) is in agreement with that in case (1)
(Fig.5).  In case (4), as the UV intensity decreases after $z=2$,
massive clouds become free from the UV background and begin to form
stars earlier than small clouds, so that the number of galaxies with
$M_{B}-5\log(h)\la -18$ increases.  Thus the effect of the UV
background remains in only small galaxies.

Next, we see the redshift evolution of luminosity functions (Fig.6),
in order to investigate the effect of the mass dependence in more
detail.  Because in this case the UV background exists before $z=5$
and is strong enough to ionize gas clouds at $z\geq 5$, the number of
faint galaxies at $z=2$ is smaller than that in cases (2) and (3).
Note that minimum mass blocks begin to collapse at $z\sim 10$ in our
model.

At $z=2$, the number of galaxies with intermediate magnitude
$M_{B}-5\log(h)\la -18$ for case (4) is in agreement with that for
case (2).  At $z=0.4$, the number of such galaxies for case (4)
increases rapidly compared with that for case (2), although the number
of faint galaxies ($M_{B}-5\log(h)\ga -18$) is nearly constant.  This
shows that galaxies with magnitude $M_{B}-5\log(h)\la -18$ become free
from the UV background and begin to form stars at $z\sim 0.4$.

Note that when the UV background vanishes completely as for case (3),
such differences depending on the mass of galaxies in the
photoionization effect do not occur.  Therefore, the effect that the
epoch when clouds become free from the UV background and begin to form
stars depends on their mass, in the case when the UV intensity
decreases continuously can be a physical mechanism that lessens the
slope at the faint end of the luminosity function, as is favoured by
the observations.

From Figs.3-6, the luminosity function in case (4) seems to be closer
to the observations than in case (1) for $0.1\la J_{-21}\la 10$ under
the assumption that our criterion of star formation is correct.

\begin{figure}
\epsfxsize=9cm
\epsfbox{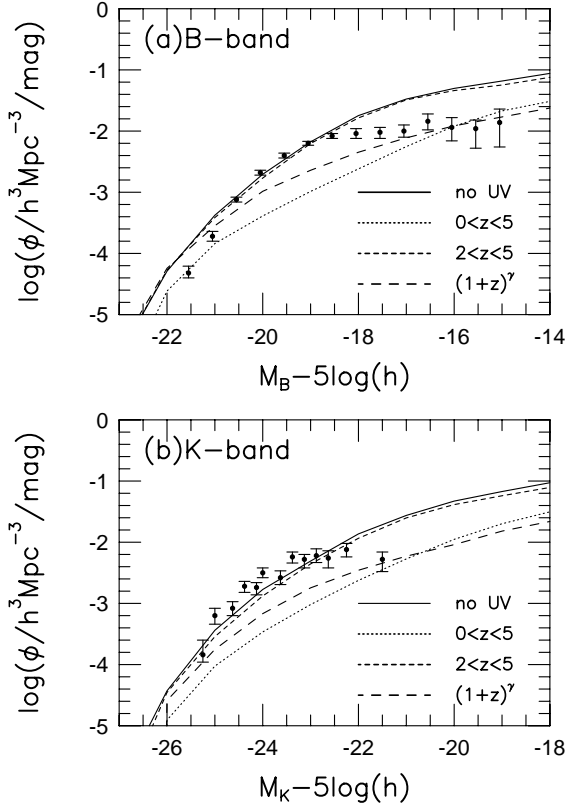}
\caption{Same as Fig.3, but for the strong condition.}
\end{figure}

\begin{figure}
\epsfxsize=7cm
\epsfbox{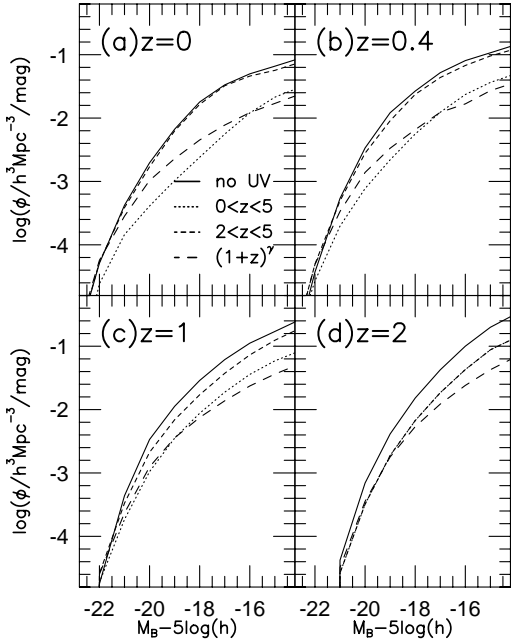}
\caption{Same as Fig.4, but for the strong condition.}
\end{figure}

\subsection{Colour distribution at $z=0$}
In Figs.7(a) and 7(b), we show the $B-K$ colour distributions of
galaxies in the weak condition for dwarf galaxies ($-19.5\leq
M_{B}\leq -17$) and for luminous galaxies ($M_{B}\leq -19.5$),
respectively.  The types of lines are the same as Fig.3.  The dotted
histograms are the observational data by Mobasher, Ellis \& Sharples
(1986), which are normalised by the total number of observed galaxies.

In the weak condition the difference between the lines is small as for
the difference between the luminosity functions for our four models
shown in Fig.3.  In case (2), the colour at the peak of the
distribution of faint galaxies becomes slightly redder, because of the
suppression of star formation at low redshifts.  On the other hand, we
cannot find any significant differences in the colour distributions of
luminous galaxies.

In Fig.8, we show the same figures as in Fig.7, but for the strong
condition.  In Fig.8(a), the colour at the peak of the distribution in
case (2) becomes redder than that in the case of no UV radiation,
because star formation at low redshifts is suppressed by the
photoionization.  When the UV radiation ceases at $z=2$ in case (3),
the colour becomes bluer than that in the case of no UV radiation,
because stars are newly formed after the UV radiation vanishes.  In
case (4), the colour distribution becomes much broader.  This is
because the epoch at which stars can be formed depends on the mass of
objects when the UV intensity continuously decreases at low redshifts,
as mentioned in the previous subsection (see Fig.1).

On the other hand, the colour distributions for luminous galaxies are
not affected so much by the UV radiation, because of the mergers of
galaxies as mentioned in the previous subsection (Fig.8b).  We find,
however, that the distributions have two peaks in cases (2) and (4).
The reason that one of the peaks is bluer than the peak of the faint
galaxies (Fig.8a) is that massive objects store hot gas which cannot
be turned into stars in the progenitors.  Such massive objects, which
are formed by mergers of the small progenitors can form stars at low
redshifts with little effect from the UV radiation.  Luminous
galaxies, the progenitors of which collapse at high redshift when the
effect of the UV background is negligible, make the other peak redder
in colour (see Section 3.1.2.1).

\begin{figure}
\epsfxsize=9cm
\epsfbox{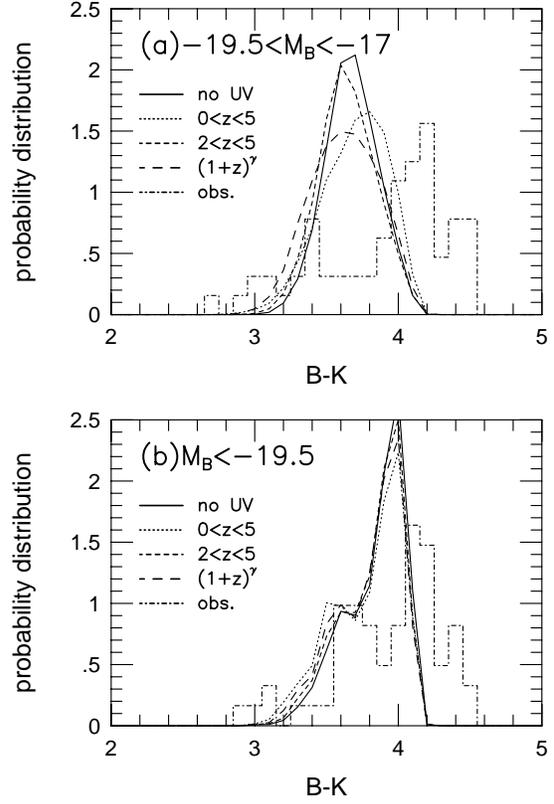}
\caption{Colour distributions in the weak condition.  The solid lines,
  dotted lines, short-dashed lines and long-dashed lines show cases
  (1) no UV, (2) UV radiation exists in $0\leq z\leq 5$, (3) $2\leq
  z\leq 5$, and (4) $J\propto (1+z)^{\gamma}, \gamma=4$ for $z\leq 2$
  and $\gamma=-1$ for $z\geq 2$.  The histograms of the dash-dotted
  lines are from Mobasher et al.(1986), which are divided by the total
  number of galaxies.}
\end{figure}
\begin{figure}
\epsfxsize=9cm
\epsfbox{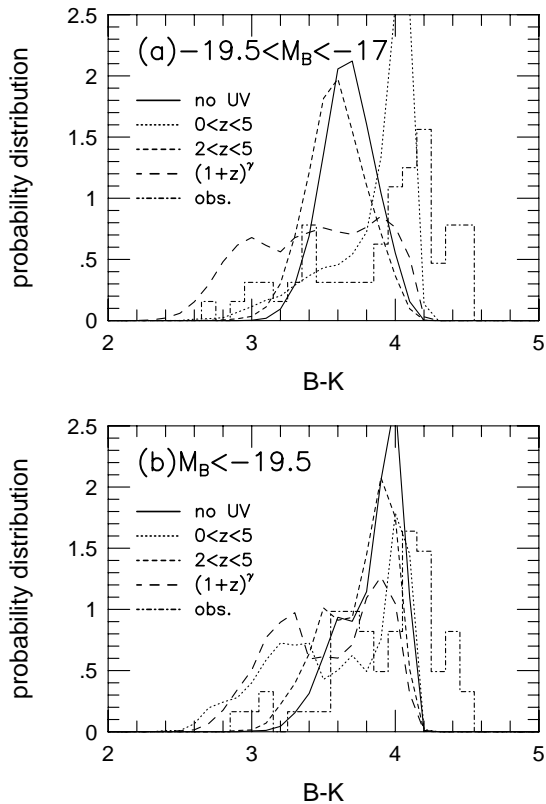}
\caption{Same as Fig.7, but for the strong condition.}
\end{figure}

\subsection{Evolution of colour distributions}
In Figs.9 (the weak condition) and 10 (the strong condition), we show
the redshift evolution of the colour distribution of galaxies in the
galaxies' rest frame for cases (1) and (4).  The solid lines, dotted
lines, short-dashed lines and long-dashed lines denote the
distributions at $z=0, 0.4, 1$ and $2$, respectively.  The thin lines
and the thick lines show case (1) no UV, and (4) continuously evolving
UV intensity.  Figs.9(a) and 10(a) show the distributions for faint
galaxies ($-19.5\leq M_{B}\leq -17$), and Figs.9(b) and 10(b) for
luminous galaxies ($M_{B}\leq - 19.5$).

In Fig.9(a), the difference between cases (1) and (4) becomes slightly
larger at lower redshifts.  As mentioned in Section 3.1.2.3, at high
redshifts $z\ga 2$, mergers of galaxies occur frequently, so that the
UV intensity becomes effectively small.  Therefore, such merging
galaxies are only slightly affected by the UV background.  Thus the
distributions at high redshifts are those for these merging galaxies.
Besides, as colours of galaxies depend not on their luminosity but on
their stellar ages, the colours of galaxies at high redshifts are
little affected by the UV background, while the number of galaxies
decreases when the UV background exists.  On the other hand, at low
redshifts, the hot gas photoionized by the UV photons at $z\sim 2$
begins to cool and form stars.  Therefore, the distributions at low
redshifts broaden to blue side, compared with case (1).  These
properties are more remarkable in the case of the strong condition
(Fig.10a).  Note that we do not consider the chemical evolution
process in this paper, so that the colour reflects the epoch of the
major star formation in each galaxy.

In Figs.9(b) and 10(b), we also see the same properties as Figs.9(a)
and 10(a).  The differences between the distributions in cases (1) and
(4) for the luminous galaxies are smaller than those for the faint
galaxies.  The distributions for faint galaxies have longer tails
towards the blue side than those for luminous galaxies, because the
fractions of young stars in such faint galaxies are larger than those
in luminous galaxies.  Thus we find that the effect of the UV
background on faint-galaxy formation remains until later epochs than
that on luminous-galaxy formation (see Section 3.1.2.3).  This also
arises from the fact that the UV background less affects massive
haloes.

\begin{figure}
\epsfxsize=9cm
\epsfbox{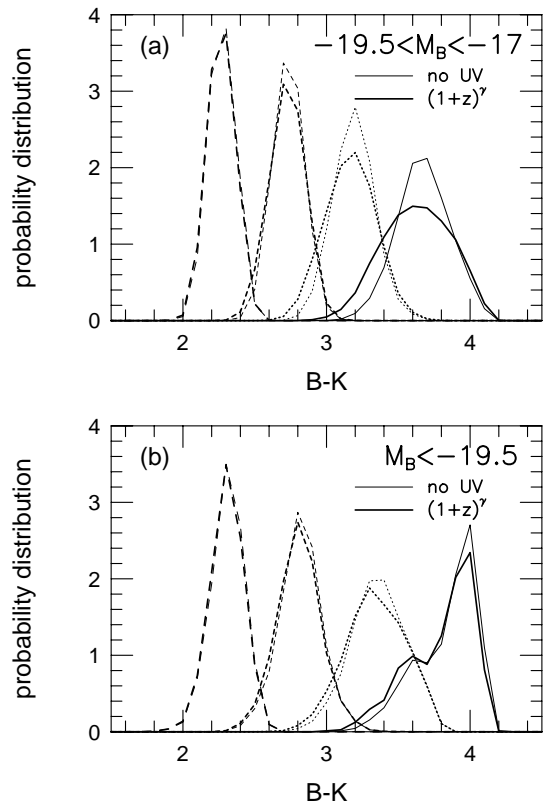}
\caption{Colour distributions at various redshifts in the galaxies' rest
  frame.  (a) Colour distributions of galaxies selected by $-19.5\leq
  M_{B}\leq -17$.  (b) Colour distributions of galaxies selected by
  $M_{B}\leq -19.5$.  The solid lines, dotted lines, short-dashed
  lines and long-dashed lines denote the distributions at $z=0, 0.4,
  1,$ and $2$, respectively.  The thin and thick lines show cases (1)
  no UV, and (4) continuously evolved UV intensity.  In all cases the
  weak condition is used.}
\end{figure}
\begin{figure}
\epsfxsize=9cm
\epsfbox{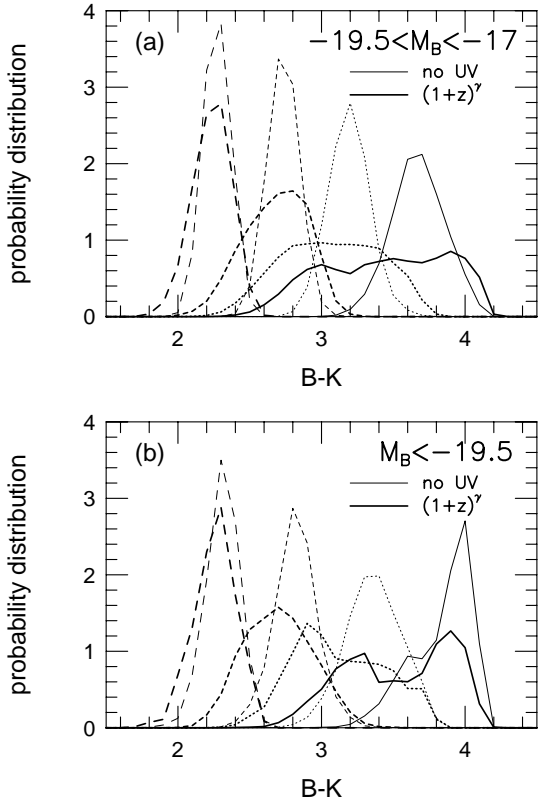}
\caption{Same as Fig.9, but for the strong condition.}
\end{figure}

\section{CONCLUSIONS AND DISCUSSION}
We investigate the effects of the UV background radiation on galaxy
formation with the semi-analytic model, which is an extension of Cole
et al. (1994) by including the photoionization process.  We find that
the UV background radiation mainly affects low-mass objects
corresponding to dwarf galaxies.

If the UV radiation field exists until the present epoch (case 2), the
number of dwarf galaxies at present becomes small.  In hierarchical
clustering scenarios, dwarf-scale galaxies are statistically formed at
high redshifts, and such galaxies are merged into luminous objects.
Therefore, dwarf galaxies at present are formed at low redshifts.
Thus, when the effect of the UV background is strong, the star
formation in such dwarf galaxies is suppressed by the UV radiation.
On the other hand, luminous galaxies are less affected, because such
luminous galaxies are formed in high overdensity regions and formed at
high redshift owing to the high overdensity environments when the
effect of the UV radiation is negligible.  This represents that it is
very important to consider mergers of galaxies.

When the UV radiation vanishes at $z=2$ (case 3), the number of dwarf
galaxies hardly changes, but the colour becomes bluer because baryonic
gas in clouds recombines and begins to form stars after the UV
radiation field vanishes.  These blue galaxies might be related to the
`{\it faint blue galaxies}' (e.g. Ellis 1997).

More realistically, in the case that the UV background evolves as
$(1+z)^{\gamma}$, where $\gamma=4$ for $z\leq 2$ and $\gamma=-1$ for
$z\geq 2$ (case 4), the slope of the luminosity functions in the faint
end becomes gentler and closer to the observed one, compared with that
in the case of no UV background.  The reason why the slope changes is
considered as follows.  When the UV intensity declines after $z=2$,
large clouds become free from the effects of the UV background at
relatively high redshifts, while small objects are affected until low
redshifts.  Therefore, the number of galaxies with intermediate
magnitudes ($M_{B}-5\log(h)\la -18$) increases, compared with the case
when the UV background exists with a constant intensity until the
present epoch.  Thus the effect of the UV background remains only in
small galaxies.  Similarly, the colour distribution becomes very broad
because the epoch at which stars can be formed in the decreasing UV
radiation field differs according to the mass of the haloes.

We also show the redshift evolutions of the luminosity functions and
the colour distributions.  If the UV intensity is sufficiently strong
($J_{-21}\ga 1$), the difference between the evolutions of the UV
background clearly appears in the difference between the evolutions of
the luminosity function.  Moreover, because the effect of the
photoionization depends on the mass of galaxies, we obtain much
information about the properties of the UV background from the colour
distributions for different classes of absolute magnitudes of
galaxies.

Although our knowledge of processes of the galaxy formation and nature
of the UV background is far from complete at present, it is certain
that, when the observations of evolutions of luminosity function and
colour distributions are realized, as well as the observations of the
evolution of Ly$\alpha$ clouds, in the near future, they will play a
significant role in investigating the evolution of the UV background
and its effects on galaxy formation.  Thus we believe that it is a
powerful tool for investigating the evolution of the UV background to
observe the redshift evolutions of the luminosity function and the
colour distribution of galaxies.

It should be noted that our conclusion is contrary to the conclusion
of Chiba \& Nath (1994).  They have reported that the number of
massive galaxies becomes small in the UV radiation field.  This
difference results from considering the mergers of galaxies.  We
emphasize that we must construct a galaxy formation model including
the mergers of galaxies, because the epoch of the formation of
luminous galaxies is not the same as the epoch of the formation of
stars in the luminous galaxies.

At the present moment, we do not know the exact criterion for forming
stars with the UV radiation.  In order to study the galaxy formations
more quantitatively, we need to investigate the criterion for forming
stars when the UV background exists in detail, and also the dependence
on the merging history model and the other parameters in the model.
We will also need to include the chemical evolution process in order
to compare our results with the observations in the near future.

\section*{ACKNOWLEDGMENTS}    
We wish to thank M. Umemura, R. Nishi, M. Chiba, K. Okoshi, H. Susa,
and Y. Tajiri for useful suggestions.  We also thank Prof. S. Charlot
for permitting his spectral evolution model.  We also thank the
anonymous referee who led us to a substantial improvement of our
paper.  This work was supported in part by Research Fellowships of the
Japan Society for the Promotion of Science for Young Scientists
(No. 2265 and No. 3167), and in part by the Grant-in-Aid for
Scientific Research (No. 10640229) from the Ministry of Education,
Science, Sports and Culture of Japan.  The calculations were performed
in part on VPP300/16R and VX/4R at the Astronomical Data Analysis
Center of the National Astronomical Observatory, Japan.

\bsp

\begin{thebibliography}{}   
\bibitem{}Babul A., Rees, M.J., 1992, MNRAS, 255, 346
\bibitem{}Bajtlik S., Duncan R.C., Ostriker J.P., 1988, ApJ, 327, 570
\bibitem{}Bechtold J., 1994, ApJS, 91, 1
\bibitem{}Carswell R.F., Morton D.C., Smith M.G., Stockton A.N.,
Turnshek D.A., Weymann R.J., 1984, ApJ, 278, 486
\bibitem{}Chiba M., Nath B.B., 1994, ApJ, 436, 618
\bibitem{}Cole S., Kaiser N., 1988, MNRAS, 233, 637
\bibitem{}Cole S., Aragon-Salamanca A., Frenk C.S., Navarro J.F., Zepf
S.E., 1994, MNRAS, 271, 781
\bibitem{}Dekel A., Rees M.J., 1987, Nat, 326, 455
\bibitem{}Efstathiou G., 1992, MNRAS, 256, 43P
\bibitem{}Ellis, R.S., 1997, ARA\&A, 35, 389
\bibitem{}Giallongo E., Cristiani S., D'Odorico S., Fontana A.,
Savaglio S., 1996, ApJ, 466, 46
\bibitem{}Gunn J.E., Peterson B.A., 1965, ApJ, 142, 1633
\bibitem{}Kauffmann G., Charlot S., 1998, MNRAS, 294, 705
\bibitem{}Kauffmann G., White S.D.M., Guiderdoni, 1993, MNRAS, 261, 921
\bibitem{}Kulkarni V.P., Fall S.M., 1993, ApJ, 413, L63
\bibitem{}Loveday J., Peterson B.A., Efstathiou G., Maddox S.J., 1992, 
ApJ, 90, 338
\bibitem{}Lu L., Sargent W.L.W., Womble D.S., Takada-Hidai M., 1996,
ApJ, 472, 509
\bibitem{}Mobasher B., Ellis R.S., Sharples R.M., 1986, MNRAS, 223, 11
\bibitem{}Mobasher B., Sharples R.M., Ellis R.S., 1993, MNRAS, 263, 560
\bibitem{}Nagashima M., Gouda N., 1997, MNRAS, 287, 515
\bibitem{}Peebles P.J.E., 1993, The Principles of Physical Cosmology.
Princeton Univ. Press, Princeton
\bibitem{}Pei Y.C., 1995, ApJ, 438, 623
\bibitem{}Roukema B.F., Peterson B.A., Quinn P.J., Rocca-Volmerange
B., 1997, MNRAS, 292, 835
\bibitem{}Somerville R.S., Primack J.R., 1998, preprint (astro-ph/9802268)
\bibitem{}Tajiri Y., Umemura M., 1998, ApJ, 502, 59
\bibitem{}Williger G.M., Baldwin J.A., Carswell R.F., Cooke A.J.,
Hazard C., Irwin M.J., McMahon R.G., Storrie-Lombardi L.J., 1994, ApJ,
428, 574
\end{thebibliography}
\end{document}